\documentclass{article}

\PassOptionsToPackage{numbers, sort&compress}{natbib}
\usepackage[preprint]{neurips_2022}

\usepackage[utf8]{inputenc} 
\usepackage[T1]{fontenc}    
\usepackage{hyperref}       
\usepackage{url}            
\usepackage{booktabs}       
\usepackage{amsfonts}       
\usepackage{nicefrac}       
\usepackage{microtype}      
\usepackage{xcolor}         
\usepackage{graphicx}
\usepackage{wrapfig}
\usepackage{amsmath}
\usepackage{enumitem}
\graphicspath{ {./figures/} }
\title{G2GT: Retrosynthesis Prediction with Graph to Graph Attention Neural Network and Self-Training}

\author{%
  Zaiyun Lin \\
    Beijing Stonewise Technology \\
  \And
  Shiqiu Yin\thanks{Correspondence to: yinshiqiu@stonewise.cn}\\
  Beijing Stonewise Technology\\
  \And
  Lei Shi\\
  Beijing Stonewise Technology
  \And 
  Wenbiao Zhou\\
  Beijing Stonewise Technology
  \And
  YingSheng Zhang\\
  Beijing Stonewise Technology
}

\begin{document}

\maketitle

\begin{abstract}
\label{abstract}
 Retrosynthesis prediction is one of the fundamental challenges in organic chemistry and related fields. The goal is to find reactants molecules that can synthesize product molecules. To solve this task, we propose a new graph-to-graph transformation model, G2GT, in which the graph encoder and graph decoder are built upon the standard transformer structure. We also show that self-training, a powerful data augmentation method that utilizes unlabeled molecule data, can significantly improve the model’s performance. Inspired by the reaction type label and ensemble learning, we proposed a novel weak ensemble method to enhance diversity. We combined beam search, nucleus, and top-k sampling methods to further improve inference diversity and proposed a simple ranking algorithm to retrieve the final top-10 results. We achieved new state-of-the-art results on both the USPTO-50K dataset, with top1 accuracy of 54\%, and the larger data set USPTO-full, with top1 accuracy of 50\%, and competitive top-10 results.
\end{abstract}

\section{Introduction}

Retrosynthesis prediction is one of the fundamental challenges in organic chemistry and related fields. The goal is to find a suitable set of reactants to synthesize product molecules. In recent decades, researchers have developed and used a variety of computational retrosynthesis tools to aid in designing synthetic routes for new molecules. The existing methods can be divided into template-based, template-free, and semi-template, and the latter two are primarily deep learning based methods.

Template-based works\cite{corey1985computer} are rule-based systems in which the target molecule is applied sequentially to all known templates, and then sets of reagents are selected based on predefined strategies. Such systems, which rely on expert knowledge, are unable to keep up with the growing number of reported reactions. Some other works\cite{law2009route,coley2017prediction} employ machine learning to extract reaction templates automatically from reaction data sets. However, this method requires atom mapping information, a scheme for mapping atoms in reactants to atoms in products. Atom mapping prediction is still an unsolved challenge \cite{chen2013automatic}, such tools rely on libraries of reaction templates and expert rules to function. Eventually, seemingly automatic techniques are still premised on reaction templates. These works share the limitations of all template-based methods.

The semi-template methods\cite{liu2020decomposing,somnath2021learning,wang2021retroprime,shi2020graph,yan2020retroxpert}.  decompose the retrosynthesis into two sub-tasks: (1) predict the reaction centers of the product and(2) Split the product into synthons based on the reaction center and convert the synthon into a complete molecule. The limitations are obvious: these methods rely on high-quality atom mapping reaction datasets to extract reaction centers. Hence, the training sets of such methods are limited to small subsets of reactions and are not scalable to real-world applications. 

The template-free methods turn one-step retrosynthesis prediction into a Seq2Seq translation task where the product SMILES \cite{weininger1988smiles} and set of reactants SMILES are the “source language” and the “target language.” Such methods can use all existing reaction data and generalize reaction patterns that have not yet been discovered. Liu et al.\cite{liu2017retrosynthetic} first adopts such framework using Long Short-Term Memory (LSTM) network and achieves performance comparable to template-based methods. Then, the work AT\cite{tetko2020state} , which adopts transformer\cite{vaswani2017attention} and utilizes SMILES augmentation strategies, brings state-of-the-art results on both USPTO-50k and the larger dataset USPTO-Full, which is a challenging dataset for other methods because it lacks reliable atom mapping information. 

Hence, inspired by template-free methods success and recent development in GNN\cite{ying2021transformers, dwivedi2020generalization,dwivedi2020benchmarking,kreuzer2021rethinking,velivckovic2017graph}, and its SOTA application on Molecule Graph Representation learning\cite{rong2020self, hu2019strategies,ying2021transformers}, we formalize the retrosynthesis problem as a Graph2Graph problem and propose the Graph2Graph Transformer(G2GT) model, which consists of the Graphormer\cite{ying2021transformers} as the graph feature encoder and a novel graph decoder. We adopt graph representation because molecules are naturally graphs, with atoms as nodes and chemical bonds as edges, making them ideal for GNN models. Besides, unlike SMILES, graph models are not affected by atom order. 

Our contribution can be summarized as follows:
\begin{enumerate}[leftmargin=*,noitemsep]
\item We propose a novel general Graph2Graph model for the one-step retrosynthesis task. Our model inherits the benefits of the template-free methods and replaces the sequence representation with graph representation. Both decoder and encoder (Graphormer) are built upon the standard transformer structure to attain the global receptive field and parallelization capability. 
\item We propose using self-training as a data augmentation strategy, combining sampling and beam search results with a frequency ranking to retrieve the top-10 results and a novel weak-ensembling method to increase the diversity. Our ablation study shows that these techniques significantly improve the model's performance.
\item G2GT advances state of the art for both USPTO-50k and USPTO-Full datasets and proved the robustness and generalization ability on a data set from a different source, Reaxys\cite{lawson2014making}. 
\end{enumerate}

\section{Preliminary}
In this section, we recap the preliminaries Transformer and Graphormer\cite{ying2021transformers}. 

\subsection{Transformer}
At the core of the transformer is the attention mechanism. We simply present a single head attention function for illustrative purposes.

$Z=\left[z_{1}^{\top}, \cdots, z_{n}^{\top}\right]^{\top} \in \mathbb{R}^{n \times d}$be the input of self-attention module where $z_i$ is the hidden representation at position $i$ and $d$ represents the hidden dimension. Then, the attention layer linearly projects $Z$ to create three learned representations: Key, Query, and Value vectors. The attention function is computed as :

\begin{equation}
\operatorname{Attention}(Q, K, V)=\operatorname{softmax}\left(\frac{Q K^{T}}{\sqrt{d_{k}}}\right) V
\end{equation}

The scalar products between Q and K can capture their similarity and represent the relationships that matter. 

\subsection{Graphormer} \label{graphormer}
Edge Encoding, Spatial Encoding, and Centrality Encoding are three Graphormer's main designs, which act as inductive biases in the neural network learning the graph representation.

\textbf{Spatial Encoding.}  Nodes in graphs are not ordered in a sequential manner, but can exist in a multidimensional spatial space and are connected by edges.
Graphormer choose $\phi(v_i,v_j)$ to be the distance of the shortest path between $v_i$ and $v_j$ if the two nodes are connected. Denote $A_{(i,j)}$ as the $(i,j)-element$ of the Query-Key product matrix A, we have $b_\phi(vi,vj )$, a learnable scalar indexed by $\phi(vi , vj )$ in Eq. \ref{GraphormerAttention}.

\textbf{Edge Feature.} For each ordered node pair $(v_i,v_j)$, it finds (one of) the shortest path $SP_{ij}=(e_1,e_2,…,e_N )$  from $v_i$ to $v_j$ , and computes an average of the dot-products of the edge feature and a learnable embedding along the path. The proposed edge encoding incorporates edge features via a bias term $c_{i,j}$ to the attention module. We have the final Equation:

\begin{equation}\label{GraphormerAttention}
A_{i j}=\frac{\left(z_{i} W_{q}\right)\left(z_{j} W_{k}\right)^{T}}{\sqrt{d}}+b_{\phi\left(v_{i}, v_{j}\right)}+c_{i,j}
\end{equation}

\textbf{Centrality Encoding.} Graphormer uses the degree centrality as an additional signal to the transformer network. The softmax attention can catch the node importance signal in the queries and keys by employing centrality encoding in the input, shown in Eq.\ref{CentralityEncoder} where $d$ is the degree of the node in the graph that corresponds to an atom's connectivity in a molecule.

\begin{equation}\label{CentralityEncoder}
z_i^{(0)}=x_i+d_{deg(v_i)}
\end{equation}

Therefore the graphormer can capture both the semantic correlation and the node importance in the attention mechanism.
\label{gen_inst}

\begin{figure}
  \centering
  \includegraphics[width=0.6\linewidth]{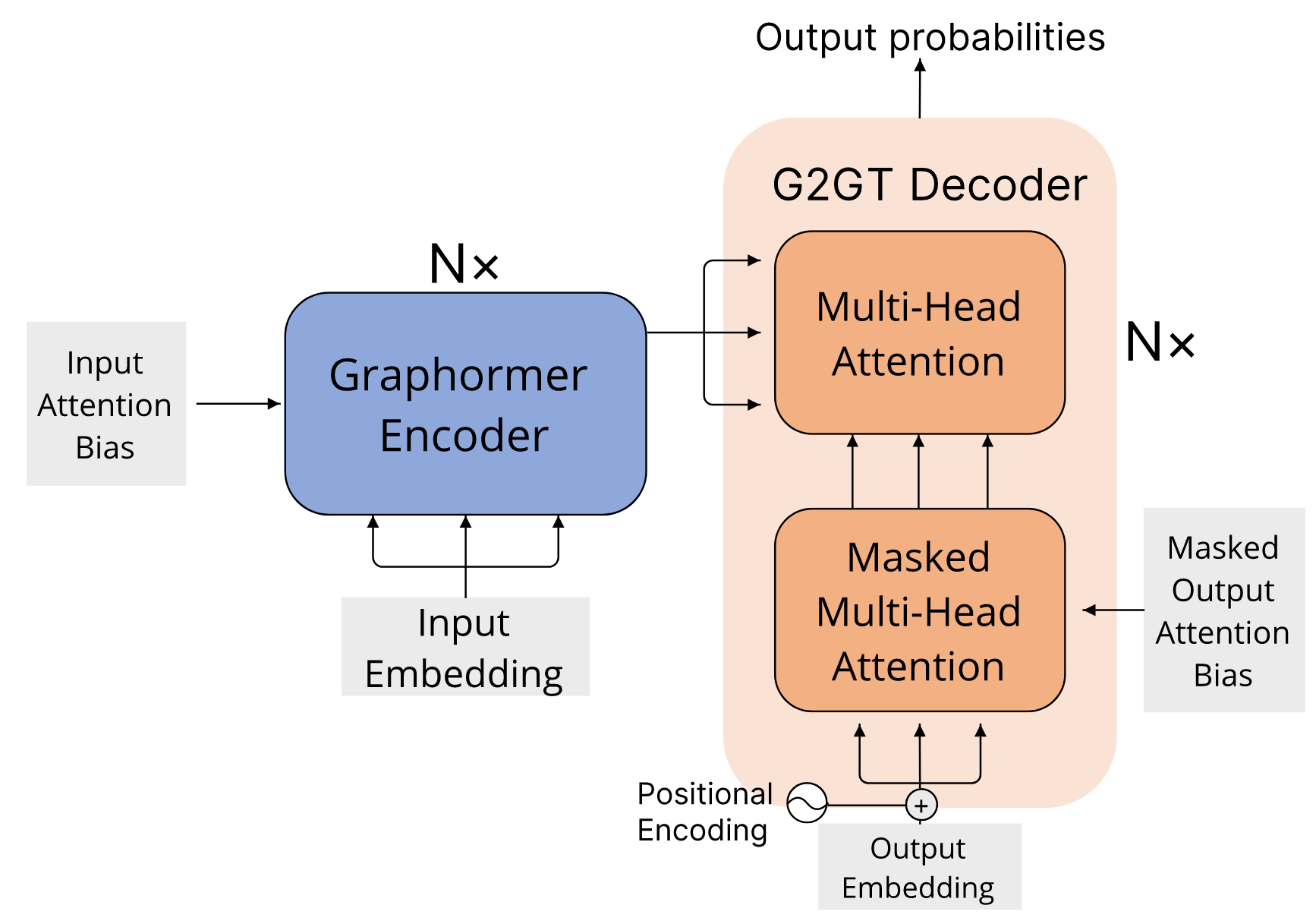}
  \caption{Overview of the G2GT model architecture.}\label{fig:ModelArchitecture}
\end{figure}

\section{The G2GT framework}

\subsection{G2GT architecture}
As discussed in the introduction, we define the reaction prediction problem as a Graph-to-Graph problem and propose a novel Graph2Graph transformer architecture G2GT. The G2GT's overall architecture are shown in Figure\ref{fig:ModelArchitecture}.

The encoder module adopted the Graphormer\cite{ying2021transformers} proposed by Ying et al., which encodes the input Graph \(G = {V, E}\) ,where \(V = {v_1, v_2, ... , v_n}\), \(n = |V|\) is the number of nodes, to a sequence of continuous representations \(z = (z_1, ..., z_n)\). Given $z$, the decoder generates an output Graph sequence \((y_1, ..., y_m)\) of symbols one element at a time. For each newly generated step, the model will calculate a new centrality and spatial information based on the updated graph state and feed to the model as decoder’s attention bias. The model then generates the next sequence based on the current graph state.

\subsection{Graphormer encoder}\label{GraphormerEncoder}
As discussed in the introduction, it is essential to incorporate the structural information of graphs into the model. We adopted Graphormer\cite{ying2021transformers} as our encoder, which is built upon the standard Transformer architecture and licensed under MIT License. It successfully leverages the structural information of graphs into the Transformer model using three simple but effective designs.


\subsection{Decoder}
\begin{figure}
  \centering
  \includegraphics[width=0.8\linewidth]{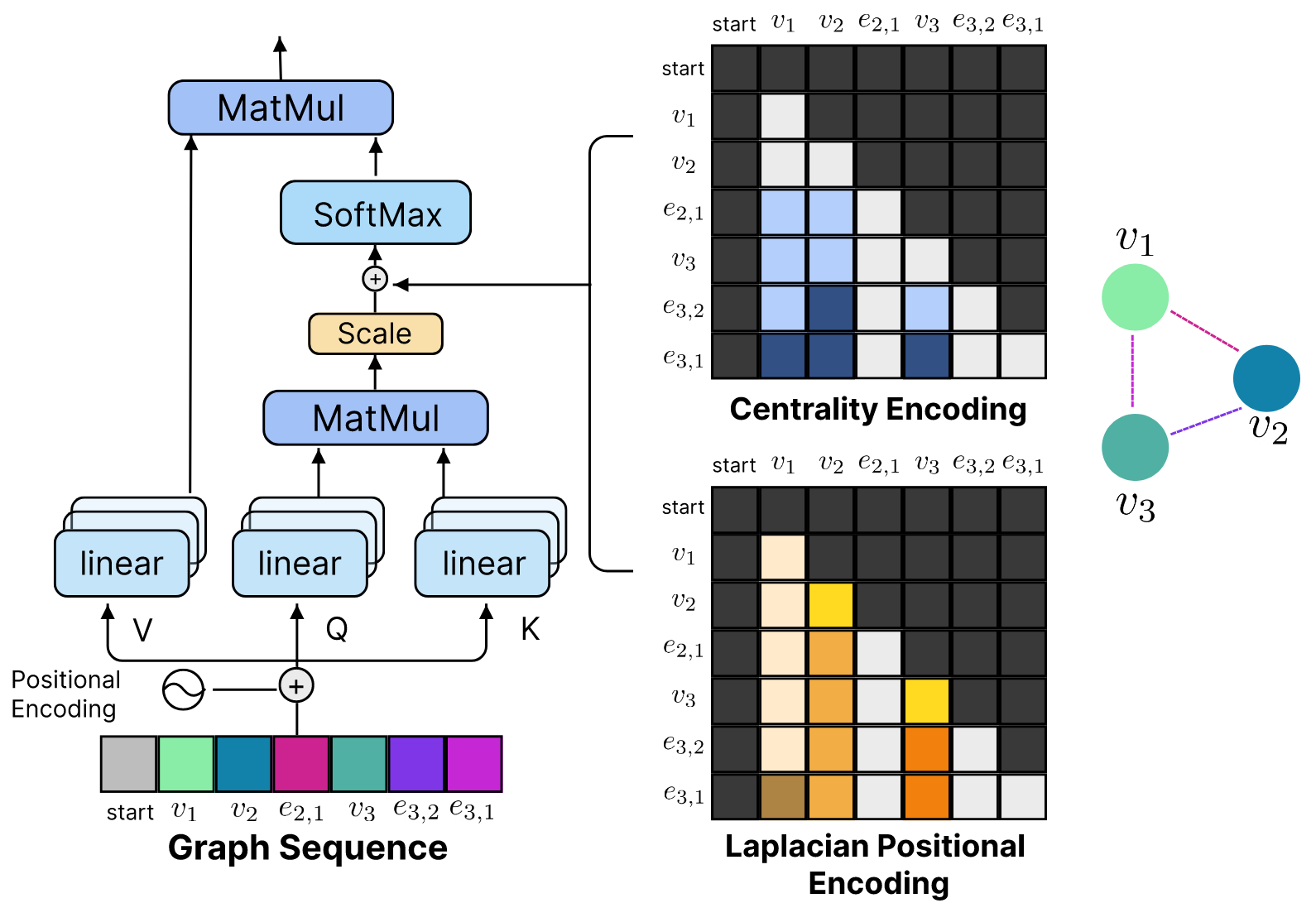}
  \caption{An illustration of G2GT's decoder including Laplacian positional encoding, and centrality encoding.}\label{fig:decoder}
\end{figure}

This section shows several critical designs in the G2GT decoder, consisting of a novel graph sequence and inductive attention bias that are compatible with the masked attention module. Furthermore, we present the implementations of our novel graph decoder. Lastly, we will discuss the advantages of our proposed model over the previous sequence-based decoder. See Figure \ref{fig:decoder} for illustration of the decoder module.

\subsubsection{Graph representation}
Once the input graph is encoded and given the input molecule representation vector z, the decoder auto-regressively infers the output graph sequences given z. The output graph sequence is defined as below:
A molecule is denoted as an undirected graph \(G={V,E}\), where V is the set of atoms and \(E\) is the set of bonds. Given a fixed atom ordering \(\pi\), the molecule graph G is uniquely represented by its weighted adjacency matrix \(A\in{R}^{(n\times n)}\), where n is the number of atoms. The weight of the edge is determined by its attributes. (e.g., Let \(e_{(i,i+1)}\) be a bond between atom \(i\) and \(i+1\) , if it is a single bond \(e_{(i,i+1)=1}\), if it is a double bond \(e_{(i,i+1)=2)}\). 

We can obtain a graph sequence \(Y={v_1,v_2,e_{2,2-1},…,v_n,e_{n,n-1},…,e_{n,1}}\)  by breaking $A$ by rows. Each atom \(v_i\) is followed by the sequence of the weighted edges ${e_{i,i-1},…,e_{i,1}}$ connecting the atom $v_i$ to its previous atom $v_{i-1}$. To compress the length of the whole sequence, we remove each atom’s trailing null edges, e.g. $\{e_{i,i-1},…, e_{i,3 },null,null\} \rightarrow \{e_{i,i-1},…, e_{i,3 }\} $. For non-trailing consecutively appearing null edges,  we compress these null edges to a single number $gapN$ that represents the number of consecutive null edges, e.g. $\{e_{i,i-1},…,null,null,e_{i,j}\} \rightarrow \{e_{i,i-1},…,gap2,e_{i,j}\}$. An illustrated example is shown in Appendix \ref{fig:GraphRep}.

\subsubsection{Centrality Encoding (Decoder)}
We can directly add the centrality encoding to our node input in the encoder stage. However, due to the information leakage problem, we cannot directly add the final centrality information to the graph sequence input during the decode stage. When the decoder auto-regressively generates the following sequence, we must prevent it from knowing future information. Furthermore, as the graph grows, a node’s centrality may increase if it connects to a new node. Therefore, to address this problem, we use the attention bias to hold the centrality information of all time steps. 

Concretely, for any graph G, we pre-calculate a centrality matrix $E \in \mathbb{R}^{n \times n \times d} $, where $n$ is the graph sequence length and total generation steps and $d$ is the number of head in the multi-head attention. Denote $e_{i,j}$, with dimension $d$, as $(i,j)$-element of $E$ and $g_i^j$ as $i$th element in the graph sequence at time step $j$ and $b$ as the learnable embedding indexed by the degree of the node $deg(g_i^j)$, we have: 
\begin{equation}\label{eq:centrality}
e_{i,j}=
\begin{cases}
    b_{deg(g_i^j)} & \text{if $g_i^j$ is a node}\\
    0 & \text{if $g_i^j$ is not a node} \\
\end{cases}
\end{equation}
See figure \ref{fig:centrality} in appendix for an illustrated example.

Denote $A_{i,j}$ as the (i, j)-element of the Query-Key product matrix $A$ of the decoder, we have:
\begin{equation} \label{eq:centrality+attention}
A_{i j}=\frac{\left(z_{i} W_{q}\right)\left(z_{j} W_{k}\right)^{T}}{\sqrt{d}}+e_{i,j}
\end{equation}

\subsubsection{Laplacian Positional Encoding (LPE)}

As discussed in the introduction, traditional methods like Liu et al.\cite{liu2017retrosynthetic} represent the graph as a sequence. However, sequence alone cannot fully capture the structural information of the graph. To preserve this information, previous studies\cite{dwivedi2020generalization, velivckovic2017graph}
have proposed to use the eigenfunctions of their Laplacian as positional encodings. 

Instead of adding the LPE directly to the node input, we use the attention bias to hold the LPE similar to how the centrality encoding is held. By concatenating the $m$-lowest eigenvalues with their associated eigenvectors, we generate an embedding matrix of size $2\times m$ for each node. $m$ is a hyper-parameter for the maximum number of eigenvectors to compute. We add masked padding for graphs where $m>N$. Then, a linear layer is applied on the dimension of size $2$ to generate new embeddings of size $k$. In this case, $k=d=$number of heads. Finally, the sequence is reduced to a fixed $k$ -dimensional node embedding via sum pooling. For any graph $G$, we pre-calculate a LPE matrix $L\in \mathbb{R}^{n \times n \times d}$. Denote $l_{i,j}$ as $(i,j)$-element of $L$, and $lpe$ as the node wise LPE, we have:
\begin{equation}
l_{i,j}=
    \begin{cases}
    lpe_{g_i^j} & \text{if $g_i^j$ is a node}\\
    0 & \text{if $g_i^j$ is not a node}
    \end{cases}
\end{equation}
Then we modify the Eq.\ref{eq:centrality+attention} further with the equation: 
\begin{equation}
    A_{i j}=\frac{\left(z_{i} W_{q}\right)\left(z_{j} W_{k}\right)^{T}}{\sqrt{d}}+e_{i,j}+l_{i,j}
\end{equation}

\begin{figure}
  \centering
  \includegraphics[width=1\linewidth]{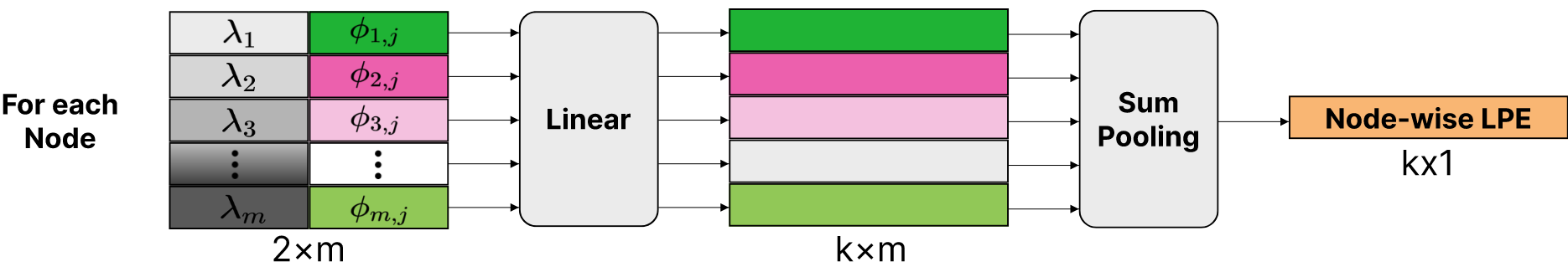}
  \caption{Laplacian Positional Encoding. $\phi_{i,j}$ denotes the eigenvector of node $j$ paired to the $i$-th lowest eigenvalue . $m$ is the hyper-parameter for the maximum number of eigenvalues and eigenvetors, and $k$ is the dimension of the LPE.}
  \label{LPE}
\end{figure}

By concatenating the eigenvalues with the normalized eigenvector, this model directly addresses problems in previous methods\cite{dwivedi2020generalization,velivckovic2017graph}. Namely, it normalizes the eigenvectors, pairs eigenvectors with their eigenvalues, and treats the number of eigenvectors as a variable. Furthermore, the model is aware of multiplicities and can linearly combine or ignore some of the repeated eigenvalues. In addition, as in other research\cite{dwivedi2020benchmarking,dwivedi2020generalization}, we randomly flip the sign of the pre-computed eigenvectors during training to increase sign ambiguity invariance. Figure \ref{LPE} shows the proposed LPE module.

\subsubsection{Positional Encoding (PE)}

In the decoder, we add the positional encodings (PE) to allow the network to learn the generation order of the graph sequence. We use sine and cosine functions of different frequencies as the positional encodings. 
\begin{equation}
 \mathrm{PE}_{(p o s, 2 i)}=\sin \left(\frac{p o s}{10000^{2 i / d_{e m b}}}\right), \mathrm{PE}_{(p o s, 2 i+1)}=\cos \left(\frac{p o s}{10000^{2 i / d_{e m b}}}\right)  
\end{equation}

\section{Data augmentation and decoding strategy }
\subsection{Self-training}
Our graph2graph model’s encoder is graph permutation invariant. So we cannot use smiles augmentation schemes proposed in previous works like Tetko et al.\cite{tetko2020state}. We used an approach called the self-training method similar to the AlphaFold\cite{jumper2021highly} to enhance our model’s generalizability. We first train a model using the training set, then the model is used to predict the reactions on the external molecule set and select the high confidence reactions to add to the training set to retrain the model. 

\subsection{Diversity}
\begin{figure}
  \centering
  \includegraphics[width=1\linewidth]{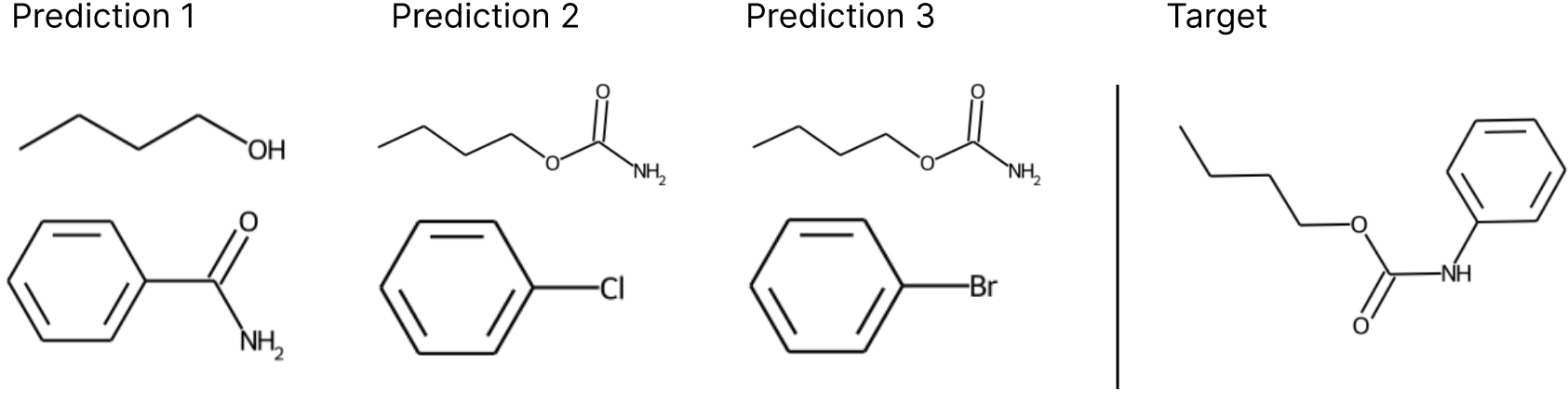}
  \caption{On the right is the input target molecule, and on the left are three possible reactant predictions. The first prediction is a Hofmann rearrangement reaction, while the second and third predictions are both Buchwald coupling reaction. The only difference between the latter two is the functional group (Cl vs Br).}
  \label{fig:diversity}
\end{figure}

Diversity is a major concern in retrosynthesis prediction. A product can go through different reaction classes and be synthesized by a diverse set of alternative reactants. The variation of the reactants' leaving groups is another case of retrosynthesis diversity. An example is shown in figure \ref{fig:diversity}. Therefore, we designed a combination of strategies to increase the prediction diversity.

\subsubsection{Beam search, Top-p, and Top-k Sampling}

During inference, we use top-$p$\cite{holtzman2019curious} (aka nucleus sampling) and top-$k$ sampling as our decoding method. Top-$p$ sampling selects tokens from the smallest accessible set whose cumulative probability exceeds the probability $p$. As a result, the size of the collection of tokens might change dynamically depending on the probability distribution of the next token. Given a distribution $P(x|x_{1:i-1})$, we define its top-p token $V(p)\subset V$ as the smallest set such that:
\begin{equation}\label{Top-p}
\sum_{x\in V(p)}^{}{P(x|x_{1:i-1})\ge p}
\end{equation}

Temperature is also used to adjust the probability. Setting $t \in [0,1)$ skews the distribution towards high probability events, which implicitly lowers the mass in the tail distribution. 
\begin{equation}\label{temperature}
p\left(x=V \iota \mid x_{1: i-1}\right)=\frac{\exp \left(u_{\iota} / t\right)}{\sum_{\iota^{\prime}} \exp \left(u_{\iota}^{\prime} / t\right)}
\end{equation}

\textbf{Frequency ranking.} To output top-n results, we can sample m times, $m>n$ . After attaining $m$ number of samples, we use the occurrence frequency as the ranking index to retrieve the top-n results. When counting the occurrence, we additionally used predictions decoded by beam search as complimentary samples.
This ranking technique is similar to the ranking strategy used in Augmented Transformer(AT)\cite{tetko2020state}. They proposed that the frequency of predicted smiles could be utilized as a confidence metric for (retro)synthesis prediction and can provide a quantitative estimation of the most likely reactions among top-n predicted outcomes. 
\subsection{Weak Ensemble }
To further increase the diversity of the predicted reactants, we explored the influence of the training set’s reaction class distribution on the predicted reactants. As shown in figure \ref{fig:diversity}, a product molecule could have multiple plausible reactants. Given such molecules, the distribution of different reaction classes in the training set will cause the model bias towards certain reaction classes.

To alleviate this problem, we proposed a novel method inspired by ensemble learning. Specifically, we randomly split the training set into $n$ sets. Then we concatenate a special node that holds a tag id information to the input of the product molecule. We jointly train all sets together. Finally, at the inference stage, we attach different tag id to a product molecule and perform sampling from these variants of the same product. Our ablation experiments \ref{tab:ablation} showed that this technique increased the top-n accuracy.

\section{Experiments}
\subsection{Data}\label{data}
We mainly use the patent mining work of Lowe et al.\cite{lowe2012extraction}, USPTO-50k, and USPTO-Full as our benchmark dataset to compare to previous work. Additionally, we used a nonpublic dataset extracted from the Reaxys database to compare the performance against a previous end-to-end transformer model. 

\textbf{50k}\cite{liu2017retrosynthetic}
We used a training set filtered from the USPTO database containing 50k reactions classified into ten reaction types. We split it into 40k, 5k, and 5k reactions for training, validation, and test sets, as proposed by Liu et al\cite{liu2017retrosynthetic}.

\textbf{Uspto-Full} \cite{dai2019retrosynthesis,tetko2020state} 
The original USPTO-full was created by Dai et al.\cite{dai2019retrosynthesis} Reactions with multiple products were duplicated into multiple ones with one product each. The authors also remove duplications in reactions and those with wrong mapping. Then we use a further filtered version developed by Tetko et al.\cite{tetko2020state}, which eliminates incorrect reactions such as those with no products or only single ions as reactants. This filtering reduced the size of the train/valid/test sets by 4\% on average to 769/96/96 k.

\subsection{Baselines \& Evaluation metric}
Baseline consist of both template-based, semi-template, and template-free methods. All the results are obtained from the orginal report, since we follow the same experiment setting.

According to the standard evaluation method used by Liu et al.\cite{liu2017retrosynthetic}, the prediction is considered correct if and only if all the reactants of a reaction are correctly predicted. We use top-k accuracy as the evaluation metric, which is commonly used in literature. Finally, to make the comparison with ground truth, we use the canonical SMILES generated by RDKit.

\begin{table}
  \caption{Top-k accuracy on USPTO-50k}
  \label{tab:50k}
  \centering

  \begin{tabular}{llcccc}
    \toprule
    Method Type & Methods &\multicolumn{4}{c}{Top-K Accuracy$(\%)$} \\\midrule
     & &k=1 &3 &5 &10\\
    \midrule
   Semi-Template & GraphRetro & $53.7$ & $68.3$ & $72.2$ & $75.5$ \\
   & G2Gs & $48.9$ & $67.6$ & $72.5$ & $75.5$ \\\midrule
    Template-based & GLN & $52.5$ & $69$ & $75.6$ & $83.7$ \\ \midrule
    Template-free  & AT & $53.2$ &  & $\mathbf{80.5}$ & $85.2$ \\
    & MEGAN & $48.1$ & $70.7$ & $78.4$ & $\mathbf{8 6 . 1}$ \\
    & G2GT(this work) & $\mathbf{54.1}$ & $69.9$ & $74.5$ & $77.7$ \\
    \bottomrule
  \end{tabular}
\end{table}

\begin{table}
  \caption{Top-k accuracy on USPTO-Full}
  \label{tab:full}
  \centering
  
  \begin{tabular}{llcccc}
    \toprule
    Method Type & Methods &\multicolumn{4}{c}{Top-K Accuracy$(\%)$} \\\midrule
     & &k=1 &2 &5 &10\\
     
    \midrule
    Template-based & GLN & $39.3$ &  &  & $63.7$ \\
    &neuralsym&$35.8$&&&$60.8$\\\midrule
    Semi-Template & RetroPrime & $44.1$ &  & $62.8$ & $68.5$ \\\midrule
    Template-free  & MEGAN & $33.6$ &  & & $63.9$ \\
    & S-Transformer & $42.9$ & $61.0$ & & $66.8$ \\
    & AT & $46.2$ & $57.2$ & & $\mathbf{73.3}$ \\
    & G2GT& $\mathbf{49.3}$  & $\mathbf{60.0}$  & $\mathbf{68.9}$ & $72.7$  \\
    \bottomrule
  \end{tabular}
\end{table}

\subsection{Settings}\label{settings}
We use same model setting for all the experiment: \textbf{G2GT}$(L=8x2,d=768)$. Both the number of attention head and the dimension of our various attention bias modules are set to 24. The maximum number of eigenvector is set to 30. We use AdamW as the optimizer, with $\epsilon=$1e-8 and $\beta_1, \beta_2 = 0.9,0.999$. The peak learning rate is set to 2.5e-4 and end learning rate to 1e-6 followed by a quadratic decay learning rate scheduler. The total training steps for USPTO-50K are 120k and for USPTO-Full are 400k. The batch size is set to 10, and we follow the practice proposed by \cite{smith2017don} to gradually increase the batch size using gradient accumulation. During inference stage, we use top-$p$ and top-$k$ sampling. $k$ is set to $5$, $p$ is set to $0.75$ and temperature is set to $6.5$. For each molecule we sample 400 times. In regard to Weak Ensemble, we assign 20 tags for USPTO-50k, and 50 tags for USPTO-Full. All models are trained on 8 NVIDIA V100 32GB GPU. It takes about 20 hours to train on USPTO-50K and 3 days to train on USPTO-Full.
\subsection{Main results}
Table \ref{tab:50k} and \ref{tab:full} summarizes the performance of G2GT against other methods. We only evaluate the reaction class unknown settings. We achieve new state of the art on both USPTO-50k and Full top-1 accuracy and a competitive top-10 accuracy on  USPTO-Full. We are primarily interested in USPTO-Full because it is a more realistic dataset in which no reliable atom-mapping information exists. This demonstrates the advantage of our method and potential applicability in reality and reveal the drawback of the template and semi-template based methods.

\subsection{Verify generalization}
\setlength\intextsep{1pt}
\begin{wraptable}{r}{5.5cm}
  \begin{tabular}{lc}

    Methods &Top-1 Accuracy$(\%)$ \\\midrule
    AT & $32.8$ \\
    G2GT & $\mathbf{33.6}$ \\
    \bottomrule
  \end{tabular}
  \caption{Verify generalization}  \label{tab:generalization}
\end{wraptable}

The above two experiments verify the performance of our model on standard data sets. However, information leakage is unavoidable because the training and test sets come from the same source, the USPTO. Therefore, we use the external dataset from Reaxys to further demonstrate generalization and robustness against previous SOTA method AT. We use USPTO-50k model and beam search to attain the top-1 results. For the AT we replicate the training process using the augmented data provided in their reports and use the same beam width to attain the top-1 results. Table \ref{tab:generalization} shows the results.

\subsection{Ablation study on USPTO-50k}
We have developed a variety of techniques to enhance the model's performance. In this section, we examine the impact of these techniques on the final results. Table \ref{tab:ablation} shows the ablation results on USPTO-50k. The results demonstrate that the proposed self-training, sampling, frequency ranking and weak ensemble effectively improve the top-1 accuracy by $6\%$ and top-10 by $13.2\%$.
\begin{table}
  \caption{Ablation study} \label{tab:ablation}
  \centering
  \begin{tabular}{lcccc}
    \toprule
    &\multicolumn{4}{c}{Top-K Accuracy$(\%)$} \\\midrule
      &k=1 &2 &5 &10\\
    \midrule
    Baseline with beam-search& $48$&$57$&$64$&$64.5$\\
    Self-Training with beam-search  & $52.6$&$63.1$&$69.8$&$69.8$\\
    Self-Training with sampling + frequency-ranking(FR) & $53.6$& $64.7$ &$73.5$&$76.9$\\
    Self-Training + weak-ensembling with sampling + FR& $54.1$&$65.2$&$74.5$&$77.7$\\

    \bottomrule
  \end{tabular}
\end{table}

\section{Conclusion and Future work}\label{conclusion}
We implement a graph decoder structure based on Transformer for the first time, which supports the parallel prediction of all time steps in the training stage. In addition, self-training is introduced into reaction prediction to increase the number of training samples. We propose various techniques to enhance the performance diversity further. We proved the superiority of template-free and graph-based methods and applicability in the real world. 

\textbf{Limitaion.}While we consistently outperform the previous methods on top-1 accuracy, we still fell short on top-10 accuracy. Furthermore, G2GT's application on large molecules is limited due to the quadratic complexity of the self-attention module and the memory consumed by the LPE module. Therefore, further research is required.

\section{Future work}
We know that the same functional group will undergo the same or similar chemical reactions regardless of the rest of the molecule's composition\cite{nic1997compendium}. It is reasonable to believe that a model that can predict the result of the chemical reaction can reasonably characterize functional groups. Besides chemical reaction, functional groups affords compound certain physical and chemical properties. Hence, this work can be viewed as a molecular representation pretraining task. We have obtained preliminary results on some molecular property prediction tasks, which will be discussed in future works.





\bibliographystyle{abbrv}
\bibliography{bibliography}
\raggedbottom
\medskip

\appendix
\pagebreak
\section{Appendix}

Optionally include extra information (complete proofs, additional experiments and plots) in the appendix.
This section will often be part of the supplemental material.
\subsection{Graph representation}

\begin{figure}[h]
  \centering
  \includegraphics[width=1\linewidth]{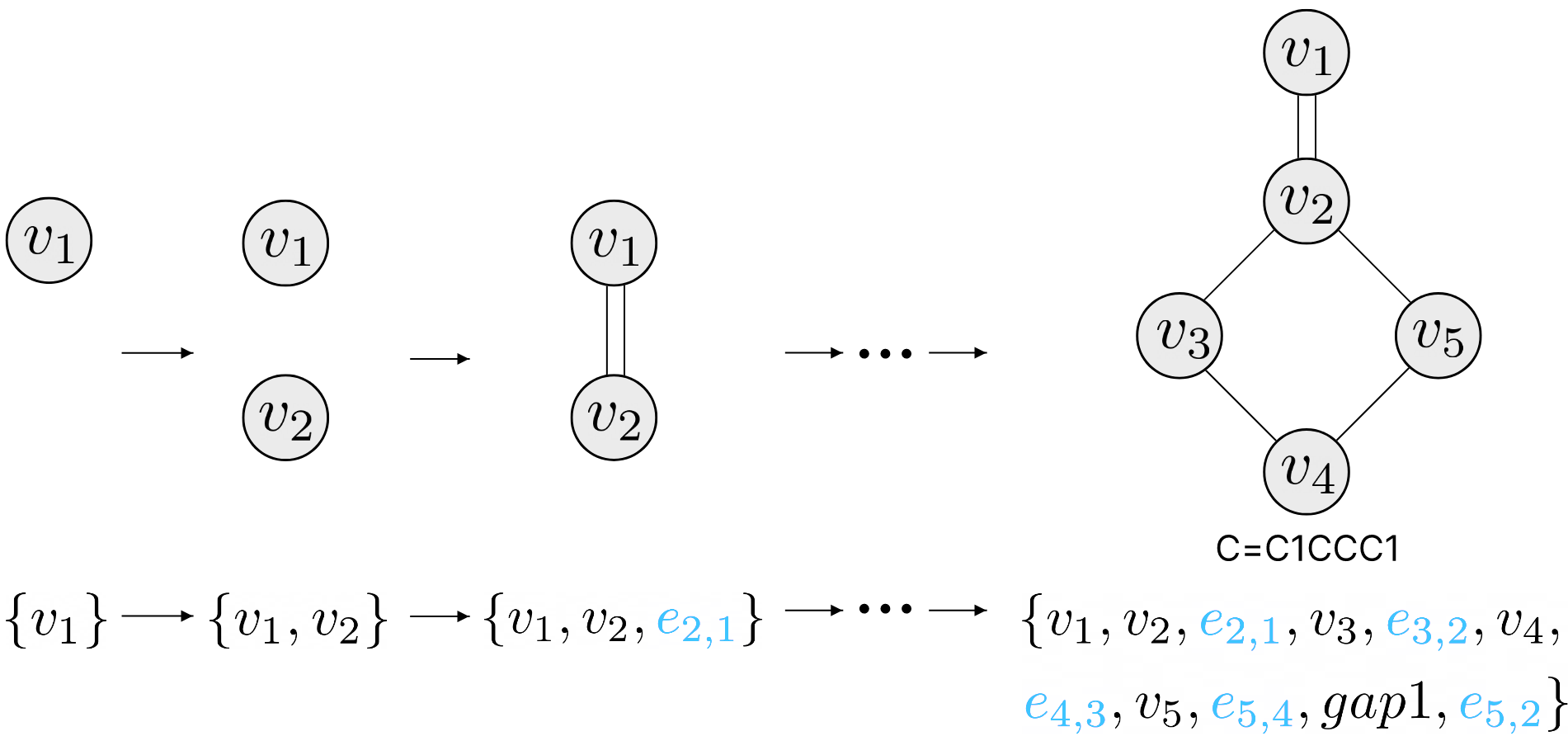}
  \caption{An example of the generation process using the proposed graph sequence representation.}
  \label{fig:GraphRep}
\end{figure}

\subsection{centrality}
\begin{figure}[h]
  \centering
  \includegraphics[width=0.8\linewidth]{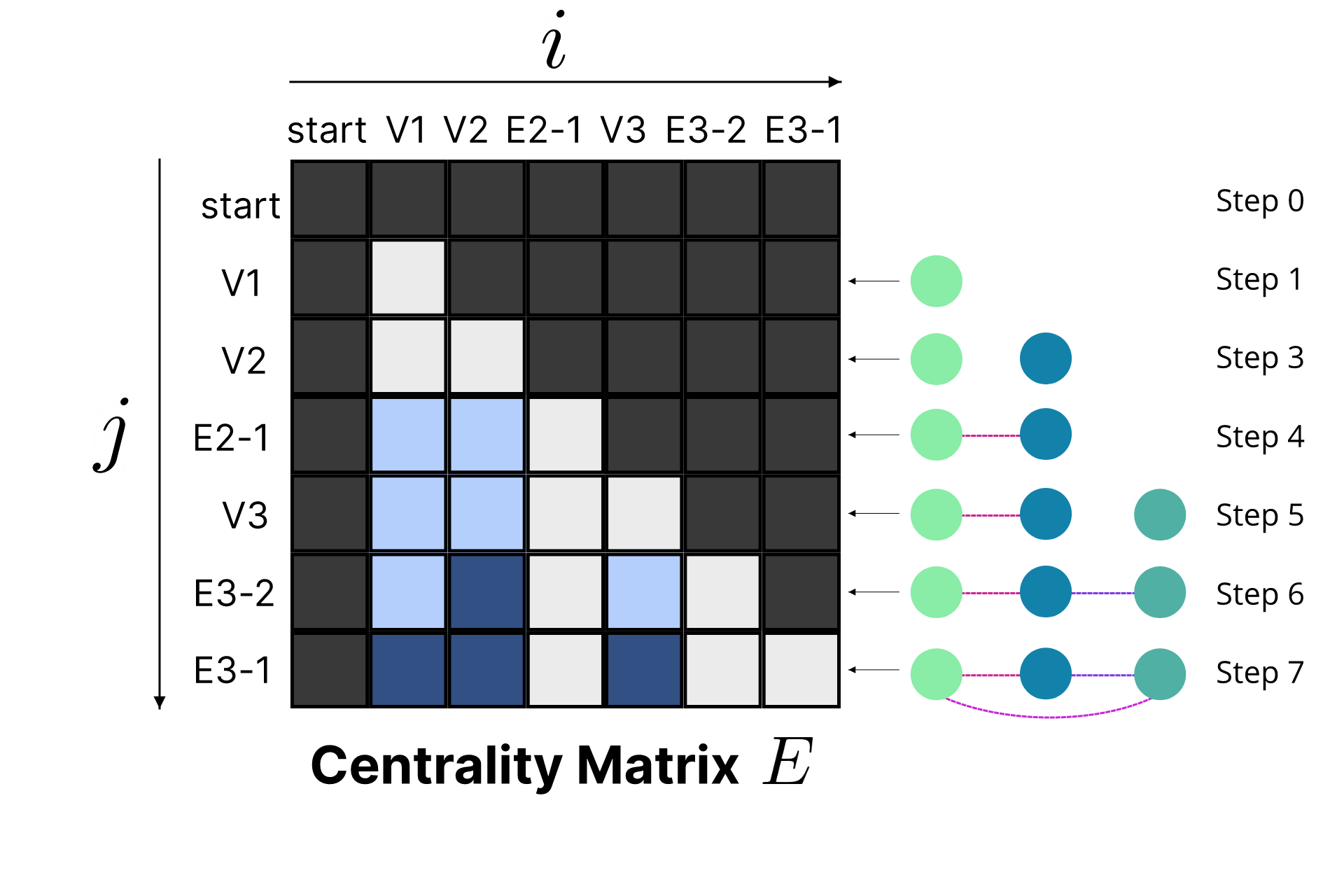}
  \caption{An example of centrality matrix given each step's corresponding graph where $i$ is the index of graph sequence and $j$ is the index of time step.}
  \label{fig:centrality}
\end{figure}

\end{document}